\documentclass[12pt,preprint]{aastex}
\usepackage{psfig}
\usepackage{amssymb}

\def\ltsima{$\; \buildrel < \over \sim \;$}
\def\simlt{\lower.5ex\hbox{\ltsima}}
\def\gtsima{$\; \buildrel > \over \sim \;$}
\def\simgt{\lower.5ex\hbox{\gtsima}}
\input{epsf.sty}
\def\eps@scaling{.95}
\def\plotone#1{\centering \leavevmode
   \epsfxsize=\eps@scaling\columnwidth \epsfbox{#1}}

\shorttitle{Bars, Bulges, AGN}
\shortauthors{Wyse}
\begin{document}
\title{On Bars, Bulges and Fuelling of Active Galactic Nuclei}

\author{Rosemary F.G.~Wyse} 
\affil{The Johns Hopkins University,
Dept.~of Physics and Astronomy, 3400 N.~Charles Street, Baltimore, MD
21218} \email{wyse@pha.jhu.edu}

\clearpage
\begin{abstract}
Both fuelling of a central black hole and the build-up of bulges are
-- at least theoretically -- facilitated by non-axisymmetric
potentials.  It is argued here that the recently established surface
brightness division between bulge-dominated galaxies that dominate the
local type 2 AGN luminosity function, and disk-dominated galaxies that
host (at best) weak AGN can be understood in terms of a simple
bar-instability criterion.

\end{abstract}


\clearpage

\section{Introduction: The Observed Surface Density Criterion}

Kauffmann et al.~(2003) have recently demonstrated, using the vast
dataset on nearby galaxies from the Sloan Digital Sky Survey, a
stellar mass, and equivalent stellar surface mass density (or surface
brightness), criterion for the separation of bulge-dominated and
disk-dominated galaxies (as classified by the value of the
concentration parameter).  They find that bulges dominate for systems
of stellar mass greater than $\sim 3 \times 10^{10}$M$_\odot$, with a
corresponding mean stellar surface density interior to the half-mass
radius of $\mu_\ast \simgt 3 \times 10^8$M$_\odot$/kpc$^2$.  Heckman
et al.~(2004) analysed the Active Galactic Nuclei (AGN) population
within the SDSS dataset, and concluded that the critical surface
density also marked the onset of AGN activity, in terms of
contributions to the volume-weighted [OIII] luminosity (their Fig.~4).
They find that significant AGN activity occurs only for $ \mu_\ast
\simgt 3 \times 10^8$M$_\odot$/kpc$^2$, peaking at $\mu_\ast \sim
10^9$M$_\odot$/kpc$^2$.

The surface density measurement used for the SDSS sample of Kauffmann
et al.~is the mean surface density within the half-light radius,
derived from a combination of model fits to the spectra to determine
the total stellar mass, and to the imaging data to determine the
Petrosian half-light radius.  The well-known `Freeman Law' (Freeman
1970) for the central surface brightness in the B-band of galactic
disks, $I_0 = 21.65$~mag/arcsec$^2$, corresponds to a central stellar
surface density of $145\, (M/L)_B$~M$_\odot$/pc$^2$ (Fall 1981).  A
thin exponential disk of scale length $\alpha^{-1}$ has the surface
density profile
\begin{equation}
\mu(R) = \mu_0 e^{-\alpha \, R} = {\alpha^2 M_D \over 2 \pi}
e^{-\alpha \, R},
\end{equation}
 where $\mu_0={\alpha^2 M_D \over 2 \pi}$ is the
central surface density.  Thus for disk-dominated galaxies, assuming an
exponential stellar disk, the half-light radius is $R_e=1.67\alpha^{-1}$  and
the mean surface density within this radius, for fixed $M/L$, is
\begin{equation}
<\mu>\,  =\,  {\frac{1}{2}M_D \over \pi R_e^2} = {\mu_0 \pi \over \alpha^2} {\alpha^2 \over \pi (1.67)^2} = {\mu_0 \over (1.67)^2} = 0.36 \mu_0. 
\end{equation}
Adopting a typical $(M/L)_B \sim 2$ from the disk models of Bell \& de
Jong (2001) gives for Freeman's Law $\mu_0 \simeq 3 \times
10^8$M$_\odot$/kpc$^2$, or $<\mu> \, \, \sim 10^8$M$_\odot$/kpc$^2$.
As shown for example by de Jong (1996), `Freeman's Law' is better
understood in terms of an upper limit for the central surface
brightnesses of disks, rather than giving the characteristic value.
The sample of de Jong (1996) has an upper limit to the B-band central
disk surface brightness of $I_0 \simlt 20$~mag/arcsec$^2$,
corresponding to $<\mu> \, \, \sim 4.6 \times 10^8$M$_\odot$/kpc$^2$.
The similarity to the SDSS-derived result is striking.

Such  an upper limit to the surface density of galactic disks has variously
been attributed in the literature to instabilities.  For example,
Dalcanton, Spergel \& Summers (1997; their Figure 5) use the
Toomre-$Q$ criterion (Toomre 1964) to provide a understanding of
`Freeman's Law'.  The Toomre-$Q$ parameter is to be evaluated locally,
as a function of radius, and involves both the local disk surface
density and disk velocity dispersion.  Mo, Mao \& White (1998; their
Figure 3),  considered the mass fraction of the disk and disk angular momentum
parameter together, and  defined a region of parameter space in which
disks are stable by the Efstathiou, Lake \& Negroponte (1982;
hereafter ELN) criterion and did not consider models that were not
stable by their criterion.  Zhang \& Wyse (2000) similarly
interpreted the ELN criterion in terms of overall baryonic mass
fraction and angular momentum parameter.

In the present paper we show that the physical conditions necessary
for the onset of bar-instability in disks (Efstathiou, Lake \&
Negroponte 1982) can be written as a pure surface density criterion.
We further show that the critical value of the surface density is very
close to that found by Kauffmann et al.~and Heckman et al.~to divide
galaxies into bulge-dominated, AGN-bright galaxies and disk-dominated,
AGN-faint galaxies.  We argue this may be understood if
non-axisymmetric perturbations such as bars are critical for efficient
fuelling of a central supermassive black hole (SMBH), and for bulge
creation.  We suggest that `classical' $R^{1/4}$ bulges may be more
associated with AGN activity than `exponential' bulges, possibly
indicating the importance of gaseous dissipation for both classical
bulge formation and AGN activity. 

\section{What turns AGN on and off?}

The AGN phenomenon depends on three schematic components (Gunn 1979)
{\it viz.} the central engine, accepted to be a massive $\simgt 
10^6 $M$_\odot$ black hole; the fuel
for the central engine, either in stellar or gaseous form; and 
lastly the mechanism by which the fuel is delivered to the
engine.  
Thus any trend in the occurence or strength of the AGN  
phenomenon should be caused by a variability in the occurence
of either the engine or the fuel, or in the efficiency of
fuelling. 
Black holes once formed are persistent,  so
it is more plausible that variations in AGN activity are 
related more to  fuelling variations  -- either in the amount of 
fuel or in the efficiency of  transportation to the black hole. 

Angular momentum provides a barrier to the inflow of fuel and   
must be lost for fuel be accreted by a central SMBH. The
importance of triaxial perturbations in the central potential well to
the efficiency of gas transportation inwards was reviewed by Shlosman, 
Begelman \& Frank (1990).  These authors envisage  that bars of gas and of
stars  form at a range of radii within a given host galaxy of an AGN
(`bars within bars'; Shlosman, Frank \& Begelman 1989), with each bar
playing a distinct role in the accretion process onto the black hole;
the maximum attainable accretion rate of gas onto a black hole can be
increased by an order of magnitude if the potential is triaxial rather
than spherical.  Similar increases in theoretical fuelling efficiency
are achieved for stellar fuel if the potential is non-axisymmetric
(Norman \& Silk 1983).

Deep imaging with the Hubble Space Telescope, particularly in the
infrared, has revealed a wealth of structure, including
bars-within-bars, in the central regions of both AGN and non-AGN host
galaxies (e.g.~Carollo et al.~2002; Erwin \& Sparke 2002; Martini et
al.~2003).  Laine et al.~(2002) found a significantly higher rate of
bars in Seyfert galaxies than in non-Seyfert galaxies, on both kpc and
sub-kpc scales.  However, as noted by those authors, even at the
H-band the effects of extinction and star formation can confuse the
identification and characterization of inner bars, so that imaging
surveys are not the ideal approach.  As suggested by the anonymous referee of this paper, 2D
spectroscopy could provide an alternative approach.

Several studies have investigated mechanisms by which a bar or
triaxial perturbation is destroyed by a central compact object,
suggesting a self-limiting process of bar fuelling and growth of a
central SMBH (e.g.~Hasan \& Norman 1990; Shen \& Sellwood 2004).
Bulges may be formed from the disrupted bars (e.g.~Norman, Sellwood \&
Hasan 1996; Pfenniger 1999); the required mass ratio of central
compact object to bar, for destruction of the bar, is of order the
mass ratio of quiescent SMBH to bulge in nearby galaxies (but depends
on several parameters, such as density e.g.~Shen \& Sellwood 2004).
Even if the central compact object does not destroy the bar, the
efficiency of fuelling may be disrupted by the changes in orbital
structure induced by scattering of stars by the central object
(e.g.~Sellwood \& Moore 1999; El-Zant et al.~2003).

But how do bars form initially? Many simulations have demonstrated
that cold, sufficiently self-gravitating disks, both stellar and
gaseous, form bars (e.g.~Ostriker \& Peebles 1973; Toomre 1977;
Efstathiou, Lake \& Negroponte 1982; Christodoulou, Shlosman \&
Tohline 1995; Bottema 2003).  Stability may be imposed either by
decreasing the level of self-gravity, for example by embedding the
disk in a dark halo, or by addition of a dense core (Toomre 1981; 
Sellwood 1989; Sellwood \& Evans 2001). 
A simple criterion for the instability
of a thin exponential stellar disk within a dark halo that provides a
flat rotation curve was derived by Efstathiou et al.~(1982; hereafter
ELN), namely bar-instability occurs if
\begin{equation}
{v_m \over (\alpha M_D G)^{1/2}} < 1.1,
\end{equation}
 where $v_m$ is the maximum rotational velocity, $\alpha^{-1}$ is the
disk exponential scale-length and $M_D$ is the total disk mass.  For a
purely gaseous disk the critical value of this ratio is somewhat
reduced, to 0.9 (Christodoulou et al.~1995).  A finite thickness,
reflecting a hotter disk, provides some stability.  As may be derived
from the simulations of Bottema (2003; his Tables 1 \& 2), disks with
an exponential scale-height up to $\sim 300$~pc, consistent with
observations, and initial gas fractions of 20\%--50\%, follow the ELN
criterion. Indeed the ELN expression has been shown to be a fairly
robust empirical criterion, irrespective of the detailed physics
driving the instability, for models of late-type (small-bulge) disk
galaxies (e.g.~Bottema 2003; O'Neill \& Dubinski 2003).  These are the
systems to which we will apply this criterion in this paper. 

The ELN criterion was re-interpreted by Mo, Mao \& White (1998) in terms
of the ratio of fractional disk-to-halo mass, $m_d$, to disk spin
momentum parameter, $\lambda_d$ such that instability occurs for
$$m_d > \lambda_d.$$ 
However, in the present context it is more
illuminating to re-write this criterion in terms of a critical surface
density, as follows.

\subsection{The disk surface density criterion for bar instability}

As noted above, an exponential disk has 
central surface density $\mu_0={\alpha^2 M_D \over 2 \pi}$. 
Upon substitution of this expression in the 
original ELN criterion, equation (3) above, one obtains bar instability for
\begin{equation}
\mu_0 > {v_m^4 \over (1.1)^4 2 \pi M_D G^2}.
\end{equation} 
The Tully-Fisher relationship can then be used to eliminate parameters
on the righthandside, assuming that the stellar mass is dominated by
the disk. As is well-known, a straightforward $L \propto M \propto
v_m^4$ relation, together with assumed virial equilibrium, immediately
gives a constant surface density (Aaronson, Huchra \& Mould, 1979,
particularly their Appendix).  Most recent investigations of the
Tully-Fisher relation however do not find a slope of 4.  Verheijen
(2001) provides correlations between $v_m$ and the K$^\prime$-band
luminosity, a good proxy for total stellar mass. From his Table 4:
$$M_K^\prime = 3.19(\pm 0.92)  - 10.5(\pm 0.4) \log (2 v_m),$$
with the units of $v_m$ here being km/s.  In what follows, we will
adopt the best-fit values of the slope and zero-point in this
relation, but one should keep the quoted uncertainties in mind.

Adopting $M_K^\prime,_\odot = +3.2 \simeq 3.19$ for the absolute
magnitude of the Sun (see Table 7, Verheijen 2001) and $(M/L)_K^\prime
= 0.5$, following the K-band models of Bell \& de Jong (2001; note
that they found only a factor of two total variation in the K-band
stellar M/L of disk galaxies across the Hubble sequence), reduces this
expression to:
$${M_D \over M_\odot} = 9.5 \left({v_m \over {\rm km/s}}\right)^{4.2}, $$
or with velocities in cm/s
$$M_D = 1.9\times 10^{13}\,\, v_m^{4.2} \qquad {\rm g}. $$ Substitution
of the resulting expression for $v_m$ into equation (4) above gives
bar instability for
\begin{equation}
\mu_0 > 5.7 M_D^{-0.04} \qquad {\rm g/cm^2}, 
\end{equation} 
and equivalently in terms of mean surface brghtness within the half-mass radius
\begin{equation}
<\mu> \, \, \,  > \, \, \,  2 M_D^{-0.04} \qquad {\rm g/cm^2}.  
\end{equation}

The mass of the disk may be eliminated from this expression by use of
the correlation between mean surface density within the half-light
radius and stellar mass found by Kauffmann et al.~(2003), again
assuming that we are in the regime where the stellar mass is dominated
by the disk.  This is:
$$<\mu> \, \, \,  \equiv \mu_\ast(M_\odot/{\rm kpc^2}) = 3 \times 10^8 \left({M_D(M_\odot) \over 3 \times 10^{10}}\right)^{0.54}.$$

Adopting this scaling between mean surface density and central surface
density provides a  mean disk surface density criterion for bar instability
of
\begin{equation}
<\mu> \, \, \,  > \, \, \,  1.7 \times 10^8 \qquad M_\odot/{\rm kpc^2}. 
\end{equation}

We have implicitly assumed purely stellar disks.  As noted above, the
ELN criterion remains valid for reasonable gas fractions, with the
total disk mass being the relevant parameter.  Given that the disk
galaxies with higher surface density tend to be earlier morphological
types, with lower gas fractions (see e.g.~de Jong 1996), the
uncertainties are within the order of magnitude estimates of this
paper. 

One might in principle use the `baryonic Tully-Fisher' relationship
(e.g.~McGaugh et al.~2000; Bell \& de Jong 2001) in place of the
stellar Tully-Fisher relation; however this would necessitate the
further assumption of equal (exponential) scale lengths for the gas
and stars to combine with the ELN criterion.  In addition, combination
with the observed stellar surface density -- stellar mass would
require knowledge of the gas/star fraction.  We have thus not
attempted this. 

\section{Discussion}

The critical mean disk surface density above, equation (7), is
remarkably close to the surface density found by Kauffmann et al.~to
characterise bulge-dominated galaxies, and by Heckman et al.~to
characterize the onset of significant contributions to the
volume-averaged Type 2 AGN luminosity function, discussed in section
1.  One expects the stellar surface density to increase once a bar
forms, since the subsequent gas fuelling to the central regions will
inevitably be accompanied by star formation, as seen both in
observations and simulations.  Thus the fact that the derived
criterion for bar instability is a factor of two or so {\it below\/} the
observationally derived surface density for bulge dominance and AGN
activity is not surprising.
Indeed, Heckman et al.~find that at higher stellar surface
densities the ratio of star-formation rate to black-hole fuelling
rate is a  constant, $\sim 10^3$, and these authors discuss how the correlation
between fossil supermassive black hole (SMBH) mass and bulge stellar
mass in nearby bulge-dominated galaxies (Gebhardt et al.~2000;
Ferrarese \& Merritt 2000) could arise from this star-formation rate
-- AGN fuelling relation.

In the context of the connections among bars, AGN and bulges, it may
be relevant that all AGN in the sample of Carollo et al.~(2002) were
hosted by a classical `$R^{1/4}$'-bulge, rather than an `exponential'
bulge (this trend was not commented on by those authors).  Exponential bulges may form from bars through secular stellar
dynamical processes (e.g.~Pfenniger 1999; Debattista et al.~2004),
while $R^{1/4}$ bulges may require efficient gaseous dissipation and
accompanying efficient star formation (e.g.~Wyse 1998; 1999).  Thus the
dissipation that is probably required to  fuel SMBH  may preferentially also lead to the formation of  $R^{1/4}$ bulges.

The picture that appears most consistent with the data is that that
bulges and AGN are causally linked with bars, resulting from
instability in high-surface density disks.  While this is not a new
picture (e.g.~Sellwood \& Moore 1999), the required elements are now
better-established.  The high-surface density disks could either form
directly from initial conditions, such as low values of the spin
parameter, $\lambda_d$ (Dalcanton, Spergel \& Summers 1997) perhaps
combined with high disk mass (Mo, Mao \& White 1998; Zhang \& Wyse
2000), or evolve towards instability through mass re-distribution, for
example by viscous processes (e.g.~Zhang \& Wyse 2000).  Exponential
bulges and $R^{1/4}$ bulges may require different amounts of gaseous
dissipation and star formation rates.  The link between AGN and
$R^{1/4}$ bulges seen in the sample of Carollo et al.~(2002) 
should be investigated with a larger
sample.  Dissipative formation of bulges and associated fuelling of a
central black hole is envisaged in the model for quasars of Silk \&
Rees (1998).  Lower luminosity AGN may form similarly.

I thank Tim Heckman for discussions.

\end{document}